\documentclass[preprint,aps,prc,showpacs,nofootinbib]{revtex4}
\usepackage{epsfig}
\usepackage{graphicx,amsmath}

\newcommand\nn{\nonumber}

\def\phi{\varphi}

\newcommand\ba{\begin{eqnarray}}
\newcommand\ea{\end{eqnarray}}
\newcommand\be{\begin{equation}}
\newcommand\ee{\end{equation}}

\begin{document}
\title{ Proton-antiproton annihilation into massive leptons }
\author{A. Dbeyssi}
\affiliation{
CNRS/IN2P3, Institut de Physique Nucl\'eaire, UMR 8608, 91405 Orsay, France} 

\author{E.~Tomasi-Gustafsson}
\email[E-mail: ]{etomasi@cea.fr}
\altaffiliation{Permanent address: \it CEA,IRFU,SPhN, Saclay, 91191 Gif-sur-Yvette Cedex, France}
\affiliation{
CNRS/IN2P3, Institut de Physique Nucl\'eaire, UMR 8608, 91405 Orsay, France} 
\author{G. I. Gakh}
%\email{gakh@kipt.kharkov.ua}
\affiliation{\it National Science Centre "Kharkov Institute of Physics and Technology"\\ 61108 Akademicheskaya 1, Kharkov,
Ukraine }
\author{M. Konchatnyi}
%\email{gakh@kipt.kharkov.ua}
\affiliation{\it National Science Centre "Kharkov Institute of Physics and Technology"\\ 61108 Akademicheskaya 1, Kharkov,
Ukraine }

%%%%%%%%%%%%%%%%%%%%%%%%%%%%%%%%%%%%%%
\begin{abstract}
%%%%%%%%%%%%%%%%%%%%%%%%%%%%%%%%%%%%%%
We extend previous calculations of polarization observables for the annihilation reaction $\bar p +p\to \ell^{-}+\ell^{+}$ to the case of heavy leptons, such as the $\tau$-lepton. We consider the case when the beam and/or the target are polarized, as well as the polarization of the outgoing leptons. We give the dependence of the unpolarized cross section, angular asymmetry, and various polarization observables on the relevant kinematical variables in the center of mass and in the laboratory system, with particular attention to the effect of the mass induced terms. 
\end{abstract}

\maketitle

%%%%%%%%%%%%%%%%%%%%%%%%%%%%%%%%%%%%%%
\section{Introduction}
%%%%%%%%%%%%%%%%%%%%%%%%%%%%%%%%%%%%%%
Reactions induced by antiproton beams will be investigated in next future at the antiproton Facility at Darmstadt, FAIR \cite{FAIR} by the  PANDA collaboration \cite{PANDA}. Among the many possible final channels in proton antiproton annihilation, we are interested here in the creation of heavy lepton ($\ell$) pairs $\bar p +p\to \ell^{-}+\ell^{+}$,  $\ell=\mu,~\tau$, through the exchange of one virtual photon of four-momentum transfer squared $q^2$. 

The case $\bar p +p\to e^{-}+e^{+}$ has been first studied in Ref. \cite{Zi62} in connection with the possibility to extract proton form factors (FFs) in time-like region, assuming one photon exchange. Polarization observables have been derived in Ref. \cite{Bi93}. More recently, single and double spin observables were derived in terms of proton FFs, and calculated on the basis of selected model which reproduce the existing data in the scattering and annihilation region \cite{TomasiGustafsson:2005kc}. Model independent expressions of single and double spin observables for the reaction $\bar p +p\to e^{-}+e^{+}$ including the contribution of two photon exchange have been given in Ref. \cite{Gakh:2005wa}. In that paper, it was already mentioned the interest in $\bar p+p$ annihilation into heavier leptons:

\begin{itemize}
\item  polarization observables corresponding to the transverse polarization of the lepton contain the factor $m_{\ell}/E$ ($m_{\ell}$ is the lepton mass, $E$ is the incident energy): in case of electron, this factor corresponds to a huge suppression, whereas, in case of $\tau$-lepton it becomes an enhancement, making the measurement easier in the GeV range. 

\item the polarization of unstable particles ($\mu $ and $\tau $) can be measured in principle through the angular distribution of their decay products. 

\item radiative corrections are different (essentially suppressed) in case of heavy leptons. 
\end{itemize}

Radiative corrections appear to be a critical issue for the extraction of proton form factors in $e+p\to e+p$ elastic scattering (see \cite{Pu10} and Refs. therein). In time-like region, all the information on the nucleon structure and on the reaction mechanism is contained in one precise measurement:  the angular distribution of one lepton at fixed $q^2$. Symmetry properties can be used to enhance or suppress C-odd terms arising from two photon exchange. The forward-backward angular asymmetry contains unique information on the ratio between the electric and magnetic FF ratio.

Following the formalism of Ref. \cite{Gakh:2005wa}, we extend the calculations of polarization observables for the annihilation reaction $\bar p +p\to \ell^{-}+\ell^{+}$, to the case of heavy leptons, such as $\mu$ or $\tau$. The main difference in the analysis of the observables in the annihilation into a $\tau$ pair, is that the $\tau$-lepton has a mass larger than the proton. Therefore, the mass can not be neglected as in the case of the annihilation into electron-positron pair. This introduces new terms in the expressions for the polarized and unpolarized observables. The aim of this work is precisely to analyze these terms.  

We consider the case when the beam and/or target are polarized, as well as the outgoing leptons. We give the dependence of the unpolarized cross section, of the angular asymmetry, and of various polarization observables on the relevant kinematical variables and study the effects induced by the heavy mass.
We derive general, model independent expressions, in terms of electromagnetic FFs in the reaction center of mass system (CMS), which is considered the natural frame for annihilation reactions, and also in the laboratory frame (Lab), since this reaction may in principle be studied at PANDA, which is a fixed target experiment.

 Some of the present results may be found in the literature. The expressions for the differential cross section and the polarization observables for the reaction $\bar p+p\to e^++e^-$ were given in terms of the nucleon electromagnetic FFs in the Lab system in the approximation of zero lepton mass in Ref. \cite{Gakh:2011mn}. In Ref. \cite{Buttimore:2006mq} the lepton mass was taken into account for the calculation of polarized and unpolarized observables in the reaction $e^++e^-\to \bar p+p$, in the reaction CMS. The observables related to the lepton polarization were not considered. Indeed, these observables may be experimentally measured, by analyzing the angular distribution of the lepton decay products. The ALEPH collaboration, for example, measured the $\tau$ polarization from  different decays such as $\tau\to\pi^- +\nu_{\tau}$, $\tau\to a_1 +\nu_{\tau}$, $\tau\to\rho +\nu_{\tau}$ \cite{Decamp:1991vz,Heister:2001uh}. 

In Section II we recall the general expressions of the lepton and hadron tensors as functions of the proton electromagnetic FFs, and give the expression for the unpolarized differential cross section in CMS and Lab systems. In Section III single, double and triple polarization observables are derived, and the effect of the mass-dependent terms is discussed. In Section IV the kinematics is illustrated and numerical results for all the observables are given, under specific assumptions for the electromagnetic FFs. Conclusions summarize the results.

%%%%%%%%%%%%%%%%%%%%%%%%%%%%%%%%%%%%%%%%%%%%%%%%%%%%%%%%%%%%%%
\section{General formalism }
%%%%%%%%%%%%%%%%%%%%%%%%%%%%%%%%%%%%%%%%%%%%%%%%%%%%%%%%%%%%%%
Let us consider the reaction:
 \be
\bar p(p_1)+p(p_2)\to \ell^{-}(k_1)+\ell^{+}(k_2),
\label{eq:eq1}
\ee
where $\ell=e$, $\mu$ or $\tau$ and the four-momenta of the particles are written in parenthesis. In the Born approximation $q=k_1+k_2=p_1+p_2$ is the four momentum of the exchanged virtual photon.

Following the formalism of Ref. \cite{Gakh:2005wa}, one can write the spin structure of the matrix element as:
\be
{\cal M}=-\frac{e^2}{q^2}j_{\mu}J_{\mu}.
\label{eq:eqM}
\ee
The leptonic and hadronic currents are:
\be
j_{\mu}=\bar u(k_1)\gamma_\mu v(k_2),
\label{eq:eqUee}
\ee
and
\be
J_{\mu}=\bar v(p_1)[G_M(q^2)\gamma_\mu+\frac{P_\mu}{M}F_2(q^2)]u(p_2),
\label{eq:eqUgg}
\ee
where $P_{\mu}=(p_1-p_2)_{\mu}/2$, and $M$ is the hadron mass. 
The quantities $G_M(q^2)$ and $F_2(q^2)$ are the magnetic and Pauli FFs of the proton. They are complex functions of the variable $q^2$. The complex nature of FFs in the time-like region of momentum transfer is due to the strong interaction in the initial state. We use below the Sachs magnetic and charge FFs, which are related to the Dirac and Pauli FFs $F_{1,2}(q^2)$ as follows:
\be
G_M(q^2)=F_1(q^2)+F_2(q^2),~G_E(q^2)=F_1(q^2)+\eta_p F_2(q^2),~ \eta_p=q^2/(4M^2).
\label{eq:eqGF}
\ee
The differential cross section is related to the matrix element squared (\ref{eq:eqM}) by 
\be
d\sigma=\displaystyle\frac{(2\pi)^4}{4{\cal I}}\overline{\left|{\cal M}\right |^2}
\displaystyle\frac{d^3\vec k_1d^3\vec k_2}{(2\pi)^6 4E_1E_2}
\delta^4(p_1+p_2-k_1-k_2), 
\label{eq:csma}
\ee
where ${\cal I}=(p_1\cdot p_2)^2-p_1^2 p_2^2$ and $E_1(E_2)$ is the energy of the $\ell^-(\ell^+)$ lepton, 
and 
\be
{\left|{\cal M}\right |^2}=\displaystyle\frac{e^4}{q^4}
L_{\mu\nu}H_{\mu\nu},~L_{\mu\nu}= j_{\mu}j_{\nu}^*,H_{\mu\nu}=J_{\mu}J_{\nu}^*.
\ee
The leptonic tensor for the case of unpolarized lepton is:
\be
L_{\mu\nu}^{(0)}= 4(k_{1\mu}k_{2\nu}+k_{1\nu}k_{2\mu})-2q^2g_{\mu\nu}.
\label{eq:eqL}
\ee
The contribution to the electron tensor corresponding to a polarized electron target is
\be
L_{\mu\nu}^{(p)}=2im_{\ell}\epsilon_{\mu\nu\alpha\beta}q_{\alpha}S_{\beta},
\label{eq:eqLp}
\ee
where $S_{\beta}$ is the polarization four-vector describing the lepton polarization.

Following Ref. \cite{Gakh:2005wa}, the hadronic tensor for unpolarized protons is:
\be
H_{\mu\nu}^{(0)}=\left ( g_{\mu\nu}-\frac{q_{\mu}q_{\nu}}{q^2}\right ) H_1+P_{\mu}P_{\nu} H_2,
\label{eq:eqW}
\ee
where 
\be 
H_1=-2q^2{\left|{G_M}\right |^2}~, 
H_2=\frac{8}{\eta_p -1}[{\left|{G_M}\right |^2}-\eta_p {\left|{G_E}\right |^2}]
.
\label{eq:eqWlh}
\ee
The matrix element squared of the reaction, $\overline{\left|{\cal M}\right |^2}$ is obtained by the contraction of leptonic and hadronic tensors (averaging over the spins of the initial particles and summing over the spins of the final particles).

\subsection{The differential cross section in CMS system}

The following analysis of the polarization observables will be done in the center of mass system (CMS). Let us define a coordinate frame (in CMS), where the z-axis is directed along the antiproton momentum $z\parallel \vec p$, the y-axis is directed along the vector $ \vec p \times\vec k$ and the $x$ axis in order to form a left handed coordinate system. In this frame, the four vectors of  the particles are
\ba
p_1&=&(E,\vec p),~ p_2=(E,-\vec p),~k_1=(E,\vec k), k_2=(E,-\vec k),~\mbox{~with~} E=\sqrt{q^2}/2, 
 \label{eq:eqp}\\
~\vec p&=&(0,0,\sqrt{\frac{q^2}{4}-M^2}),
~\vec k=
\left (\sqrt{\frac{q^2}{4}-m_{\ell}^2}\sin\theta,0,\sqrt{\frac{q^2}{4}-m_{\ell}^2}\cos\theta \right ), 
~\vec k\cdot\vec p={|{\vec k}|}{|{\vec p}|}\cos\theta , 
\nn
\ea
and $\theta$ is the CM angle of the negative lepton with respect to the antiproton beam.

The differential cross section in the reaction CMS takes the form:
\be
\frac{d\sigma}{d\Omega}^C=\displaystyle\frac{\alpha^2}{4q^6}\displaystyle\frac{\beta_\ell}{\beta_p}
L_{\mu\nu}H_{\mu\nu},
\label{eq:eqscms}
\ee
where $\beta_{\ell}^2=1-4m_{\ell}^2/q^2$ is the velocity squared of the lepton $\ell$ of mass $m_\ell$ ( $\ell=e$, $ \mu$ or $\tau$) and  $\beta_{p}^2=1-4M^2/q^2$ is the antiproton velocity squared.

In CMS, the unpolarized differential cross section is :         
\ba
\displaystyle\frac{d\sigma^C_{0}}{d\Omega}&
=&\displaystyle\frac{\alpha^2}{4q^2}\displaystyle\frac{\beta_\ell}{\beta_p}~{\cal D}^C;\nn\\
\cal{D}^C&=&\displaystyle\frac{|{G_E}|^2}{\eta_p}
(1-\beta_\ell^2\cos^2\theta                                                                  )+|{G_M} |^2 (2-\beta_\ell^2\sin^2\theta).
\label{eq:eqS}
\ea
Integrating the differential cross section (\ref{eq:eqS}) over the solid angle one finds the expression for the total cross section:
 \be
\sigma=\displaystyle\frac{\pi\alpha^2}{3q^2}
\displaystyle\frac{\beta_{\ell}}{\beta_p}
\left(2+\displaystyle\frac{1}{\eta_{\ell}}\right)\left [  \displaystyle\frac{\left|{G_E}\right|^2}{\eta_p}+ 2 
\left|{G_M}\right |^2\right ],~\eta_{\ell}=\displaystyle\frac{q^2}{4m_{\ell}^2},
\label{eq:eqSS}
\ee
which depends on the moduli squared of FFs, and does not contain any interference term. In the limit of zero lepton mass, this expression 
coincides with the results previously obtained {\cite{Zi62}:
\ba
\frac{d\sigma}{d\Omega}(\bar p p\to e^+e^-)&=&
\displaystyle\frac{\alpha^2}{4q^2\beta_p}\left [
\displaystyle\frac{|{G_E}|^2}{\eta_p}\sin^2\theta +|{G_M} |^2 (1+\cos^2\theta)\right ],
\label{eq:eqS0}\\
\sigma(\bar p p\to e^+e^-)&=&\displaystyle\frac{2\pi\alpha^2}{3q^2\beta_p}
\left(\displaystyle\frac{|{G_E}|^2}{\eta_p}
 +2|{G_M}|^2\right ).
\label{eq:eqSS0} 
\ea
From the comparison between Eqs. (\ref{eq:eqS}) and (\ref{eq:eqS0}) one can see that the terms due to the lepton mass does not change the even nature of the differential cross section with respect to $\cos\theta$, as expected from the one photon exchange mechanism, but changes the ratio of the cross section at $\theta=0{^\circ}$ or $180{^\circ}$ with respect to the cross section at $\theta=90{^\circ}$ degrees. 

%In other words, the relative contribution of the electric FF is not strongly suppressed by the factor $\eta_p$ (at the reaction threshold for $\tau$ lepton production $\eta_p\sim 4$, to be compared to $\bar p p$ annihilation into $e$ or $\mu$ pair, where $\eta_p \sim 1$). 
One can express the differential cross section as a function of an angular asymmetry ${\cal A}$, defined from the slope of the linear $\cos^2\theta$ dependence. The cross section, Eq. (\ref{eq:eqS}), can be rewritten as:
\be
\displaystyle\frac{d\sigma}{d\Omega}
=\left (\displaystyle\frac{d\sigma}{d\Omega}\right )_{\pi/2}
(1+{\cal A} \cos^2\theta),
\label{eq:eqSA}
\ee
where
\be
\left (\displaystyle\frac{d\sigma}{d\Omega}\right )_{\pi/2}= \displaystyle\frac{\alpha^2}{4q^2}\displaystyle\frac{\beta_e}{\beta_p}~
\left [\displaystyle\frac{|{G_E}|^2}{\eta_p}+|{G_M} |^2 (2-\beta_e^2)\right ],~
\label{eq:eqS90}\ee
and 
\be
{\cal A}= \beta^2_{\ell}\displaystyle\frac{\eta_p |{G_M} |^2-|{G_E}|^2}
{\eta_p  |{G_M} |^2(2-\beta^2_{\ell})+ |{G_E}|^2}.
\label{eq:eqA}
\ee
The measurement of the asymmetry ${\cal A}$ allows to determine the ratio of the moduli of the FFs through the relation:
\be
\left |\displaystyle\frac{G_E}{G_M}\right |^2=
\eta_p\displaystyle\frac{\beta^2_{\ell}-(2-\beta^2_{\ell}){\cal A}}
{\beta^2_\ell+{\cal A}}.
\ee
%%%%%%%%%%%%%%%%%%%%%%%%%%%%%%%%%%%%%%%%%%%
\subsection{The  differential cross section in Lab system}
%%%%%%%%%%%%%%%%%%%%%%%%%%%%%%%%%%%%%%%%%%%
In case of the laboratory system, the four vectors of the particles are:
$p_1=(E,\vec p)$, $p_2=(M,0)$, $k_1=(E_1,\vec k_1)$, $k_2=(E_2,\vec k_2)$. The differential cross section takes the form:
\be
\frac{d\sigma_0}{d\Omega}^L=\displaystyle\frac{\alpha^2}{4q^4}\displaystyle\frac{k^2}{Mp}(Wk-E_1p\cos\theta)^{-1}
L_{\mu\nu}H_{\mu\nu},
\label{eq:eqlab}
\ee
where $W=E+M$ is the total energy of the reaction, $E(E_1)$ and $p(k_1)$ is the energy and the magnitude of the momentum of the antiproton beam (scattered $\ell^-$ lepton), $\theta_1$ is the angle between the momenta of the antiproton beam and of the scattered $\ell^-$  in the Lab system. For simplicity, we denote $k=|\vec k_1|$ and $p=|\vec p|$, and $|G_{E,M}(q^2)|^2\equiv |G_{E,M}|^2$.

In terms of FFs, the differential cross section is written as:
\ba
\displaystyle\frac{d\sigma^L_0}{d\Omega}&=&
\displaystyle\frac{\alpha^2}{2M^2W^2}
\displaystyle\frac{(E_1^2-m_{\ell}^2)}{p(E-M)}
(Wk-E_1p\cos\theta_1)^{-1}~{\cal D}^L,\label{eq:eqSl}\\
{\cal D}^L&=&2M^2(2E_1^2-2WE_1+MW)[\eta_p|G_M|^2-|G_E|^2] +p^2(m_\ell^2+MW)|G_M|^2.
\nn
\ea
%%%%%%%%%%%%%%%%%%%%%%%%%%%%%%%%%%%%%%%%%%%%%%%%%%
\section{ Polarization Observables }
%%%%%%%%%%%%%%%%%%%%%%%%%%%%%%%%%%%%%%%%%%%%%%%%%%

The unpolarized cross section contains only the moduli squared of the form factors. In the time-like region, FFs are complex functions, due to unitarity and analyticity. The investigation of reaction (\ref{eq:eq1}) with polarized antiproton beam and/or polarized proton target carries information about the phase difference of the nucleon FFs, $\Phi=\Phi_M-\Phi_E$, where $\Phi_{M,E}=arg G_{M,E}$. This phase difference contains important information on the nucleon FFs and its determination represents a stringent test of nucleon models. 

The calculation of polarization observables requires to define a coordinate frame. Let us define a coordinate frame in the same way in Lab and CMS systems: the $z$ axis is directed along the antiproton momentum $\vec p$, the $y$ axis is directed along the vector $\vec p\times\vec k$, and the $x$ axis in order to form a left handed coordinate system.

%%%%%%%%%%%%%%%%%%%%%%%%%%%%%%%%%%%%%%%%%%%%%%%%%%
\subsection{ Single spin observables: the analyzing power }
%%%%%%%%%%%%%%%%%%%%%%%%%%%%%%%%%%%%%%%%%%%%%%%%%%

Unlike elastic $e^-p$ scattering in one-photon exchange approximation, the hadronic tensor in the reaction (\ref{eq:eq1}) contains an antisymmetric part due to the fact that nucleon FFs are complex functions \cite{Gakh:2005wa}. Therefore, in the present case, the polarization of the antiproton may lead to nonzero spin asymmetry.

The polarization four-vector of a relativistic particle of mass $M$, energy $E$ and momentum $\vec p$ , is defined by :
$$\vec s= \vec \chi +\frac{\vec p \cdot\vec \chi \vec p}{M(E+M)}, s^0=\frac{\vec p \cdot\vec \chi}{M},$$
where $\vec \chi$ is the polarization vector in the rest frame of the particle.

The hadronic and the leptonic tensors can be written as a sum of unpolarized and polarized terms. As a starting point, we consider the case when only the antiproton beam is polarized, the hadronic current can be written as: 
\be
H_{\mu\nu}=H_{\mu\nu}^{(0)}+H_{\mu\nu}^{(1)}(s_1),
\label{eq:eqD1}
\ee
where $s_{1\mu}$  is the polarization four-vector describing the antiproton polarization. The explicit expression of the tensor $ H_{\mu\nu}(s_1)$ in terms of nucleon electromagnetic FFs is:
\ba
H_{\mu\nu}^{(1)}(s_1)&=&-
\displaystyle\frac{2i}{M}
[ M^2\left |G_M\right |^2 <\mu\nu qs_1>
\nn\\
&& +(\eta_p-1)^{-1} Re G_M(G_E-G_M)^*(<\mu p_1p_2s_1>P_{\nu } -
<\nu p_1p_2s_1> P_{\mu} ) ]
\nn\\
&&+\displaystyle\frac{2}{M(\eta_p-1)}  ImG_MG_E^*(< \mu p_1p_2s_1> P_{\nu} + 
<\nu p_1p_2s_1> P_{\mu})  .
\label{eq:eqDDD}
\ea
The contraction of the spin-independent leptonic tensor $L_{\mu\nu}^{(0)}$ and the part of the hadronic tensor (\ref{eq:eqDDD}), $ L_{\mu\nu}^{(0)}H_{\mu\nu}^{(1)}(s_1)$ leads to the following expression of the differential cross section
\be
\displaystyle\frac{d\sigma^{C,L}}{d\Omega}=
\displaystyle\frac{d\sigma^{C,L}_{0}}{d\Omega}
(1+A_y^{C,L}\chi_{1y}),
\label{eq:eqscl}
\ee
where $A_y^{C,L}$ is the single spin asymmetry due to the antiproton polarization, $\vec\chi_1$ is the polarization of the antiproton in its rest frame.
The asymmetry $A_y^{C,L}$ has the form:
\be
{\cal D}^C A_y^C=\frac{\beta^2_\ell\sin 2\theta}{\sqrt{\eta_p}}Im G_MG_E^* .
\label{eq:eqayC}
\ee
One can see that taking into account the mass of the lepton leads to a factor $\beta_\ell^2$ which decreases when the antiproton energy increases.
if we take the limit of $A_y$ when the mass of leptons tend to zero, we obtain the known expression for the asymmetry of electron production \cite{TomasiGustafsson:2005kc}:
 \be
 {\cal D}^C A_y^C(m_\ell \to 0)=\frac{\sin 2\theta}{\sqrt{\eta_p}}Im G_MG_E^* ,
 \label{eqc} 
 \ee
 where
 \be
  {\cal D}^C(m_\ell \to 0)=\frac{\left |{G_E}\right |^2}{\eta_p}\sin^2\theta +{\left|{G_M}\right |^2}(1+\cos^2\theta).
 \label{eqd} 
 \ee
In the laboratory system the asymmetry can be written as:
\be
{\cal D}^L A_y^L=2M(2E_1-W)pk\sin \theta_1 Im G_MG_E^*.
\label{eq:eqayL}
\ee 
The dependence on the lepton mass is hidden in the expression for the lepton energy $E_1$. This asymmetry is determined by the component of the polarization vector which is perpendicular to the reaction plane. One can see from Eqs. (\ref{eq:eqayC}) and (\ref{eq:eqayL}) that this asymmetry vanishes in collinear kinematics, for $\theta=0{^\circ}$ or $180{^\circ}$. It can be explained, since in a parity conserving electromagnetic interaction the spin asymmetry  is determined by a correlation of the type $\vec\chi_1\cdot(\vec p\times\vec k_1)$. Therefore, it vanishes when the lepton momentum is parallel or antiparallel to the antiproton momentum. The single-spin asymmetry, although it is a T-odd observable, does not vanish in one photon exchange approximation, due to the complex nature of FFs in time-like region. This is a principal difference from the elastic $ep$ scattering, where nucleon FFs are real functions.

The measurement of this asymmetry allows to determine the phase difference of the nucleon FFs, when the moduli are determined from the unpolarized differential cross section measurement \cite{Gakh:2005wa}.

The single spin asymmetry due to the polarization of the $\tau$-lepton vanishes, since the symmetric spin-independent part of the hadronic tensor is contracted with the antisymmetric part of the spin-dependent leptonic tensor. Contrary to the spin-dependent hadronic tensor, the  spin-dependent leptonic tensor does not contain any symmetric part (over the indices $\mu$ and $\nu$). This is due to the fact that we assume that the electromagnetic interaction of the $\tau$-lepton is point-like (does not contain FFs) as in case of $e$ or $\mu$. The measurement of this asymmetry constitute an experimental test of the point-like nature of the $\tau$-lepton, at large values of $q^2$.

Note that the inclusion the two-photon exchange mechanism may lead to a non-zero value of the single-spin asymmetry due to the $\tau$-lepton polarization \cite{Gakh:2005wa}. In the case of $e$ or $\mu$ pair production, the contribution of two-photon exchange is suppressed by the presence of a factor $m_{\ell}/M$ in this asymmetry. But, in case of $\tau$ pair production, this factor will enhance the terms due to two-photon exchange. Therefore, the measurement of the polarization of a single $\tau$ lepton in the collision of unpolarized particles is a direct test of the presence of the two-photon exchange mechanism. An advantage of this measurement is that the polarization of unstable particles can be measured through the angular distribution of their decay products.

%%%%%%%%%%%%%%%%%%%%%%%%%%%%%%%%%%%%%%%%%%%%%%%%%%%%%%%%%%%
\subsection{Double spin polarization observables }
%%%%%%%%%%%%%%%%%%%%%%%%%%%%%%%%%%%%%%%%%%%%%%%%%%%%%%%%%%%

\subsubsection{Polarization transfer coefficients}

Let us consider now the polarization transfer when the antiproton beam is polarized and the polarization of the produced negative lepton is measured. 

We denote the direction of the lepton polarization vector (in its rest system)  by the indices: $\ell$ (longitudinal) along its momentum, $t$ (transverse) which is orthogonal to the momentum in the reaction plane and $n$ (normal) which is perpendicular to the  reaction plane. Then the following three independent polarization four vectors describe the lepton polarization in the reaction CMS 
\be
s_{\ell}^C= \displaystyle\frac{1}{m_{\ell}}(k,E\sin\theta,0,E\cos\theta),~
s_{t}^C= (0,\cos\theta,0,-\sin\theta),~
s_{n}^C=(0,0,1,0),
~\label{eq:eql1}
\ee
where $E(k)$ is the energy (magnitude of the momentum) of the lepton in the reaction CMS. 

The non-vanishing transfer polarization coefficients are:
\ba
T_{\ell x}^C&=&\frac{2\sin\theta}{\sqrt{\eta_p}{\cal D}^C} Re G_M G_E^*,~
T_{\ell z}^C=2\frac{\cos\theta}{{\cal D}^C}{\left|{G_M}\right |^2},
T_{n y}^C=2\displaystyle\frac{m_{\ell}}{M}\displaystyle\frac{Re G_E G_M^*}{  \eta_p {\cal D}^C},\nn\\
T_{t x}^C&=&2\displaystyle\frac{m_{\ell}}{M}\displaystyle\frac{\cos\theta}{\eta_p {\cal D}^C}Re G_E G_M^*,~
T_{t z}^C=-2\displaystyle\frac{m_{\ell}}{M}\displaystyle\frac{\sin\theta}{\sqrt\eta_p {\cal D}^C} \left|{G_M}\right |^2.
\label{eq:eqHZ}
\ea
These coefficients are $T$-even observables, and they do not vanish in the one-photon exchange approximation as well as in the elastic lepton-nucleon scattering. The coefficients $T_{n y}^C$, $T_{t x}^C$, $T_{t z}^C$ are proportional to the mass of the produced lepton and they are suppressed by the factor $m_{\ell}/M$ for $\ell=e$ or $\mu$. In the case of $\tau$-lepton this factor constitutes an enhancement of $\sim 2$. The polarization observables $T_{\ell z}^C$ and $T_{t z}^C$ are determined by the magnetic FF only, whereas 
$T_{\ell x}^C$, $T_{t x}^C$ and $T_{n y}^C$  by the factor $Re G_M G_E^*$. Therefore, the measurement of the coefficients $T_{\ell x}^C$, $T_{n y}^C$, and $T_{t x}^C$ can give in principle the information on the phase difference of the nucleon FFs. In the limit of zero lepton mass, the expressions (\ref{eq:eqHZ}) coincide with the corresponding results of Ref. \cite{TomasiGustafsson:2005kc} and of Ref. \cite{Gakh:2005wa}, neglecting the two-photon contribution.

In the Lab system, the polarization transfer coefficients have the following form:
\ba
 {\cal D}^L T_{\ell x}^L&=&2M WE_1(E-M)\sin \theta_1 Re G_MG_E^* ,\nn\\
 {\cal D}^L T_{\ell z}^L&=&2M \frac{Wp}{k} [( \eta_p-1) m_\ell^2+E_1(E_1-\eta_pM^2)]
 | G_M|^2 ,\nn\\ 
 {\cal D}^L T_{t x}^L&=&2W m_\ell M \frac{p}{k}(E_1-M) Re G_MG_E^* ,\nn\\
 {\cal D}^L T_{t z}^L&=&-m_\ell W p^2 \sin \theta_1 |G_M|^2 ,\nn\\
 {\cal D}^L T_{n y}^L&=&2  m_\ell M p^2 Re G_MG_E^* .
 \label{eq:eqt} 
 \ea
The angular dependence of the observables is hindered in $E_1$ and ${\cal D}$.

%%%%%%%%%%%%%%%%%%%%%%%%%%%%%%%%%%%%%%%%%%%%%%%%%%%%%%%%%%%%%%%%%%%%%%%%%%%%%
\subsubsection{Analyzing powers in polarized proton-antiproton collisions }
%%%%%%%%%%%%%%%%%%%%%%%%%%%%%%%%%%%%%%%%%%%%%%%%%%%%%%%%%%%%%%%%%%%%%%%%%%%%%
Let us consider the case when the polarized antiproton beam annihilates with a polarized proton target.

The part of the differential cross section which depends on the polarization of the antiproton beam and proton target can be written as:
\be
\displaystyle\frac{d\sigma^{C,L}}{d\Omega}=
\displaystyle\frac{d\sigma^{C,L}_{0}}{d\Omega}
(1+A_{ij}^{C,L}\chi_{1i}\chi_{2j}),
\label{eq:eqacl}
\ee
where $\vec\chi_2$ is the polarization vector of the proton in its rest frame and 
$A_{ij}^{C,L}$ are the spin correlation coefficients which have the following form in the reaction CMS:
\ba
{\cal D}^C A^C_{xx}&=&\sin^2\theta \beta^2_\ell\left  (\displaystyle\frac{\left|{G_E}\right
|^2}{\eta_p}+\left|{G_M}\right |^2 \right )+\displaystyle\frac{\left|{G_E}\right
|^2} {\eta_p\eta_{\ell}}, \nn\\
{\cal D}^C A^C_{yy}&=&\sin^2\theta \beta^2_\ell\left  (\displaystyle\frac{\left|{G_E}\right
|^2}{\eta_p}-\left|{G_M}\right |^2 \right )+\displaystyle\frac{\left|{G_E}\right
|^2} {\eta_p\eta_{\ell}}, \nn\\
{\cal D}^C A^C_{zz}&=&-\sin^2 \theta \beta^2_\ell \left (\displaystyle\frac{\left|{G_E}\right
|^2}{\eta_p}+\left|{G_M}\right |^2 \right )-\displaystyle\frac{\left|{G_E}\right
|^2} {\eta_p\eta_{\ell}} +2 \left|{G_M}\right|^2,\nn \\
{\cal D}^C A^C_{xz}&=&{\cal D}^C A^c_{zx}= \frac{\sin 2\theta }{\sqrt{\eta_p}}\beta^2_\ell ReG_MG_E^*.
\label{eq:e4}
\ea
One can see that at small angles the contribution which is proportional to $|G_E|^2$ dominates in the analyzing powers $A^C_{xx}$ and $A^C_{yy}$ and this effect arises from the heavy lepton mass. So, this effect is absent in the production of $e$ and $\mu$ pairs. The measurement of these observables represent a potential interest since in all observables considered above the contribution related to the electric FF is suppressed by the factor $\eta_{\ell}^{-1}$. The coefficient $A^C_{xz}$ gives information about the relative phase, through the term $\cos\Phi$. Combining this coefficient and the single spin asymmetry $A_y^C$, one can obtain an useful relation between these quantities:
\be
\tan\Phi=\frac{A_y^C}{A^C_{xz}}.
\label{eq:eqph}
\ee
The measurement of the spin correlation coefficients $A^C_{xx}$ and $A^C_{yy}$ allows to determine the ratio of the FFs moduli through the relation:
\be
\left |\displaystyle\frac{G_E}{G_M}\right | =
\displaystyle\frac{\eta_p\beta^2_\ell }
{1 +\eta_{\ell}^{-1}\cot^2\theta }
\displaystyle\frac{{\cal R}+1}{{\cal R}-1}
,~{\cal R}=\displaystyle\frac{A_{xx}^C}{A^C_{yy}}.
\label{eq:eqphr}
\ee
Note that the sum of the double analyzing powers $A_{xx}^C$ and $A_{yy}^C$ is proportional to $|G_E|^2$:
\be
 {\cal D}^C(A_{xx}^C + A_{yy}^C)=\displaystyle\frac{2}{\eta_p}\left (\displaystyle\frac{1}{\eta_\ell}+\beta^2_\ell\sin^2\theta \right )|G_E|^2.
\label{eq:eqaxy}
\ee
As it is shown below, the numerical values of these observables are large, therefore, the measurement of this sum can be considered, in principle, as a good method for the determination of $|G_E|$.
The advantage of measuring the ratio of polarization observables instead of the unpolarized cross section, is that systematic errors associated with the measurement essentially cancel as well as radiative corrections, at least the multiplicative ones, allowing more precise measurements. This is a well known general fact, and it is at the basis of the successful application of the polarization method \cite{Re68,Re74} in the space-like region (Ref. \cite{Pu10}).  

The double spin analyzing powers have the following form in the Lab system:
\ba
{\cal D}^L A^L_{xx}=&&-MW \left [2(\eta_p-1)m_{\ell}^2\left|{G_M}\right |^2
+(2E_1^2-2WE_1+MW)\left (\frac{1}{\eta_p} \left|{G_E}\right |^2 +\left|{G_M}\right |^2 \right )\right ],
 \nn \\
{\cal D}^L A^L_{yy}=&& -MW\left [ (2E_1^2-2WE_1+MW)(\frac{1}{\eta_p} \left|{G_E}\right |^2-\left|{G_M}\right |^2) -2 (\eta_p -1 )m_{\ell}^2\left|{G_M}\right |^2 \right ],
\nn \\
{\cal D}^L A^L_{zz}=&& 2M^2\left [ (2E_1^2-2WE_1+MW)\left (\left|{G_E}\right |^2+\eta_p\left|{G_M}\right |^2\right) +2\eta_p (\eta_p-1)(m_{\ell}^2+MW)\left|{G_M}\right |^2\right ], 
\nn \\
{\cal D}^L A^L_{xz}=&&{\cal D}^L A^L_{zx}= 2Mpk(2E_1-W) \sin \theta ReG_MG_E^*.
\label{eq:All}
\ea
In the limit of zero lepton mass, these double analyzing powers coincide with those obtained (in Lab system) in Ref. \cite{Buttimore:2006mq} (a part of the missing factor $\sin\theta$ in $A^L_{xz}$ and $A^L_{zx}$). On can verify that the relation (\ref{eq:eqph}) between the phase difference and the single and double analyzing powers is still valid in the Lab system. The ratio of the nucleon FFs $|G_E/G_M|$ can also be determined from the quantities $A^L_{xx}$ $A^L_{yy}$, but the corresponding expression (analogue to Eq. (\ref{eq:eqphr}) is more lengthy.
  
%%%%%%%%%%%%%%%%%%%%%%%%%%%%%%%%%%%%%%
\subsubsection{Correlation coefficients: polarized lepton-antilepton pair}
%%%%%%%%%%%%%%%%%%%%%%%%%%%%%%%%%%%%%%
Let us consider the case when both leptons are polarized in the annihilation of unpolarized antiproton and proton. The leptonic tensor can be written as: 
\ba
L_{\mu\nu}(s_1,s_2)=&& -(k_{1\mu}k_{2\nu}+k_{1\nu}k_{2\mu}-\frac{q^2}{2}g_{\mu\nu})s_1\cdot s_2+(k_{1\mu}s_{2\nu}+k_{1\nu}s_{2\mu})k_2\cdot s_1\nn\\
&&+ (k_{2\mu}s_{1\nu}+k_{2\nu}s_{1\mu})k_1\cdot s_2-(s_{1\mu}s_{2\nu}+s_{1\nu}s_{2\mu})\frac{q^2}{2}-g_{\mu\nu}k_1\cdot s_2k_2\cdot s_1,
\label{eq:eqdfh}
\ea
where $s_{1\mu} (s_{2\mu})$ is the polarization four vector describing the polarization of the $\ell^-(\ell^+)$ lepton, with $k_1\cdot s_1=k_2\cdot s_2=0$.
The non-zero correlation coefficients between the two polarized leptons, can be obtained by the contraction of the leptonic tensor with the spin-independent hadronic tensor.

The part of the cross section which depends on the polarizations of the produced leptons can be written as:
\be
\displaystyle\frac{d\sigma^{C,L}}{d\Omega}=
\displaystyle\frac{d\sigma^{C,L}_{0}}{d\Omega}
(1+C_{ij}^{C,L}\xi_{1i}\xi_{2j}),
\label{eq:eqccl}
\ee
where $\vec\xi_{1}$ ($\vec\xi_{2}$) is the polarization vector of the lepton $\ell^- (\ell^+)$  in its rest frame and $C_{ij}^{C,L}$ are the polarization correlation coefficients.
In this case, the indices $\ell,t,n$ have the following meaning in the reaction CMS: $\ell$ (longitudinal polarization) means that the polarization vectors of the negative lepton ($\vec \xi_1$) and  positive lepton ($\vec \xi_2$) in their rest frames are directed  along the momentum of the negative lepton, $t$ means that both  polarization vectors are orthogonal to this momentum (transverse polarization) and $n$ that both polarization vectors are normal to the reaction plane (normal polarization).
  
Then, the nonzero polarization correlation coefficients, in the reaction CMS 
have the following form
\ba
{\cal{D}^C}C^C_{nn}&=&\sin^2\theta \beta^2_\ell\left  [\displaystyle\frac{\left|{G_E}\right
|^2}{\eta_p}-\left|{G_M}\right |^2 \right ]+\displaystyle\frac{\left|{G_E}\right
|^2} {\eta_p\eta_{\ell}}, \nn\\
{\cal{D}^C}C^C_{tt}&=&\sin^2\theta \left (1+\displaystyle\frac{1}{\eta_{\ell}}\right )
 \left (\left|{G_M}\right |^2-\displaystyle\frac{\left|{G_E}\right|^2}{\eta_p}\right )
+\displaystyle\frac{\left|{G_E}\right
|^2} {\eta_p\eta_{\ell}}, \nn\\
{\cal{D}^C} C^C_{ll}&=&
 \sin^2\theta\left (1+\displaystyle\frac{1}{\eta_{\ell}}\right )
  \left (\displaystyle\frac{\left|{G_E}\right|^2}{\eta_p}-
 \left| G_M \right |^2 \right )+2\left|{G_M}\right |^2-\displaystyle\frac{\left|{G_E}\right
|^2} {\eta_p\eta_{\ell}},                                                                                                 \nn\\
{\cal{D}^C}C^C_{\ell t}&=&C^C_{t\ell }=\displaystyle\frac{\sin 2\theta }{\sqrt{\eta_\ell}}\left  (\displaystyle\frac{\left|{G_E}\right|^2}{\eta_p}-\left|{G_M}\right |^2 \right ).
\label{eq:eqC}
\ea

%\ba
%${\cal{D}^C}C^C_{nn}&=&{\sin^2\theta}\left  [\frac{\left|{G_E}\right
%|^2}{\eta_p}\left(1+\frac{\cot^2\theta}{\eta_\ell}\right )-\left|{G_M} \right %|^2 \beta^2_\ell \right ],\nn\\
%%
%{\cal{D}^C}C^C_{tt}&=&{\sin^2\theta}\left  [\frac{\left|{G_E}\right
%|^2}{\eta_p}\left(\frac{\cot^2\theta}{\eta_{\ell}}-1\right )+\left|{G_M}\right %|^2 \frac{\eta_{\ell} +1}{\eta_{\ell}} \right ],\nn\\
%%
%{\cal{D}^C} C^C_{ll}&=&
% \sin^2\theta \left [\frac{\left|{G_E}\right|^2}{\eta_p}
% \left (1-\frac{\cot^2\theta}{\eta_\ell}\right )+
% \left| G_M \right |^2 \left (1+2\cot^2\theta-\frac{1}{\eta_{\ell}} \right ) \right ],\nn\\
% %
%{\cal{D}^C}C^C_{\ell t}&=&C^C_{t\ell }=\frac{\sin 2\theta %}{\sqrt{\eta_\ell}}\left  %(\frac{\left|{G_E}\right|^2}{\eta_p}-\left|{G_M}\right |^2 \right ).
%\label{eq:eqdfhfth}
%\ea

From these expressions one can see that, for the $\tau$- lepton, the large mass lead to an increase of the $|G_E|^2$ term in the angular regions $\theta\sim 0^\circ$ and  $\theta\sim 180^\circ$. This effect essentially decreases for $e$ and $\mu$. One can see also that these correlation coefficients do not contain information about the phase difference of nucleon FFs.  
The polarization correlation coefficients have the following form in the Lab system:
\ba
{\cal D}^LC^L_{nn}&=&q^2\left [ (E_1^2-WE_1+\eta_p M^2)(|G_M|^2-\frac{1}{\eta_p} |G_E|^2)+(\eta_p - 1)m^2_{\ell} |G_M|^2\right ],\label{eq:eqccll}\\
{\cal D}^LC^L_{t\ell}&=&m_{\ell}q^2\frac{p}{k_2}\sin\theta \left [ (\eta_pM-E_1)\frac{1}{\eta_p}
|G_E|^2-\eta_p(E-E_1)|G_M|^2\right ],\nn\\
{\cal D}^L C^L_{\ell t} &=&m_{\ell}q^2\frac{p}{k_2}\sin\theta \left [ (\eta_pM-E_1)\frac{1}{\eta_p}
|G_E|^2-\eta_p(M-E_1)|G_M|^2\right ],\nn\\
{\cal D}^L C^L_{t t} &=&m_{\ell}\frac{q^2}{kk_2} \left \{ 
(WE_1-k^2)(k^2-WE_1+\eta_p m^2_{\ell} +\eta_p M^2) |G_M|^2 \right .\nn\\
&& \left .
-\frac{1}{\eta_p}|G_E|^2\left [\eta_p ME_1(E_1-M)(2E_1-W)-
\eta_p(\eta_p-1)M^2k^2-( E_1-\eta_p M)^2k^2\right ]\right \},\nn\\
{\cal D}^L C^L_{\ell \ell}&=& \frac{4M^2}{kk_2}\left \{\left [ \eta_p(\eta_p-1)m^2_{\ell}M^2 +
(WE_1-k^2)(E_1-\eta_p M)^2\right ] (|G_E|^2-\eta_p|G_M|^2)\right . \nn\\
&& \left . - \eta_p(\eta_p-1)\left [M^2E_1(W-E_1)(|G_E|^2+\eta_p|G_M|^2)+m^2_{\ell} 
(WE_1-W^2-k^2)|G_M|^2 \right ])\right \},
\nn
\ea
where $k_2$ is the magnitude of the $\tau^+$ lepton momentum, $k_2^2=(W-E_1)^2-m_{\ell}^2$. Note that in Lab system, the notation of polarization vectors $\ell$, $t$, $n$ refers to the direction of the momentum of the corresponding lepton.
%%%%%%%%%%%%%%%%%%%%%%%%%%%%%%%%%%%%%%
\section{Triple spin polarization observables}
%%%%%%%%%%%%%%%%%%%%%%%%%%%%%%%%%%%%%%
Let us give here some examples of triple-spin polarization observables, calculated in the reaction CMS.

We consider now the polarization observables of the production of a polarized (negative) lepton in the annihilation of polarized proton-antiproton. These coefficients are called $M_{i0jk}$ in the notations from Ref. \cite{By76}. Here the four subscripts denote, respectively, the detected particle, the associated particle, the projectile, and the target. The indices $i$, $j$ correspond to $n$, $t$, $\ell$, according to the direction of the polarization vectors of the lepton, or to $x$, $y$, $z$, referring to the direction of the hadron.

For the case of longitudinally polarized lepton, the nonzero coefficients are :
\ba
{\cal D^C}M_{\ell 0zy}={\cal D^C}M_{\ell 0yz}=&&\frac{\sin\theta}{\sqrt{\eta_p}}ImG_MG_E^*.
\label{eq:l0zy}
\ea
For the transverse and normal polarizations of the lepton, we have :
\ba
{\cal D^C}M_{t0zy}=&&{\cal D^C}M_{t0yz} =\frac{m_\ell}{M}\frac{cos\theta}{\eta_p }Im G_MG_E^* ,\nn\\
{\cal D^C}M_{n0xz}=&&{\cal D^C}M_{n0zx}=-\frac{1}{\eta_p }\frac{m_\ell}{M}Im G_MG_E^* .
\label{eq:e6}
\ea

These observables do not contain additional information about nucleon FFs. In the limit of zero lepton mass only $M_{\ell 0zy}$ and $M_{\ell 0yz}$ are nonzero, and their expressions coincide with the results obtained for the coefficients $D_{zy}$ and $D_{yz}$ obtained in Ref. \cite{Gakh:2005wa}
for the case of polarized proton-antiproton pair and longitudinally polarized electrons.

The polarization observables in the case of annihilation of a polarized antiproton beam with unpolarized proton target, when the polarization of both  lepton and antilepton is measured can be written as:
\ba
{\cal D^C}C_{tty0}&=&-\frac{\sin 2\theta}{2\sqrt{\eta_p}}\frac{\eta_\ell +1}{\eta_\ell} ImG_MG_E^*,\nn\\
{\cal D^C}C_{\ell\ell y0}&=&\frac{\sin 2\theta}{2\sqrt{\eta_p}}\frac{\eta_\ell +1}{\eta_\ell} ImG_MG_E^*,\nn\\
{\cal D^C}C_{nny0}&=&\frac{\sin 2\theta }{2\sqrt{\eta_p}}\beta_\ell^2 ImG_MG_E^*,\nn\\
{\cal D^C}C_{\ell nx0}&=&{\cal D^C}C_{n\ell x0}=-\frac{\cos\theta}{\sqrt{\eta_p \eta_\ell}}ImG_MG_E^*,\nn\\
{\cal D^C}C_{tnx0}&=&{\cal D^C}C_{ntx0}=\frac{\sin\theta}{\sqrt{\eta_p} }ImG_MG_E^*,\nn\\
{\cal D^C}C_{\ell ty0}&=&
{\cal D^C}C_{t\ell y0}=
\frac{\cos2\theta}{\sqrt{\eta_\ell \eta_p}}ImG_MG_E^*. 
\label{eq:eqch}
\ea
The remaining coefficients vanish. In the case of zero lepton mass, one finds $C_{\ell nx0}=C_{n\ell x0}=C_{\ell ty0}=C_{t\ell y0}=0$. 

Note that the ratio of any pair of these triple spin polarization observables does not depend on the nucleon FFs, but is function only of kinematics variables. This is a consequence of the one-photon exchange approximation. In case of presence of two-photon exchange, this property does not hold anymore. Therefore, in principle, the measurement of these observables gives an indication on the presence of additional mechanisms beyond Born approximation.
%%%%%%%%%%%%%%%%%%%%%%%%%%%%%%%%%%%%%
%%%results
%%%%%%%%%%%%%%%%%%%%%% 
\section{Numerical results}
\subsection{Kinematics}

In CMS, the lepton pair is emitted back to back and each lepton carries half of the total energy.
In Lab system, the kinematics for a massive lepton, in particular for $\tau$-lepton, which mass is larger than the proton mass, is essentially different from the case when the lepton mass is neglected. In case of $e$ or $\mu$, there is no limitation for the angular region of the produced (negative) lepton in the Lab system, and there is a unique relation between the energy and the angle:
\be
E_1\simeq\frac{MW}{W-p\cos\theta_1}.
\label{eq:eqe}
\ee 
When the mass of the lepton exceeds the proton mass, there is a maximum limiting angle for the lepton emission $\theta_{max}$, which depends on the lepton mass and on the incident energy:
\be
\cos\theta_{max}=\displaystyle\frac{W\sqrt{m_{\ell}^2-M^2}}{m_\ell p},
\label{Eq:thmax}
\ee
and it is illustrated at vertical lines in Fig. \ref{Fig:Eth}.

In Lab system, from the conservation laws of energy and momentum one finds that one angle corresponds to two possible values for the energy of the emitted $\tau^-$ lepton:
\be
E_1^{\pm}=\displaystyle\frac
{MW^2\pm \sqrt{p^2\cos^2\theta_1\left [W^2(M^2-m_{\ell}^2)+m_{\ell}^2p^2\cos^2\theta_1 \right ]}}
{(W^2-p^2\cos^2\theta_1)}.
\label{eq:en}
\ee
This is illustrated in Fig. \ref{Fig:Eth}, for three incident energies $E=6.85$ GeV,  just above threshold (blue dash-dotted line), $E=15$  GeV (black solid line) and $E=30$ GeV  (red dashed line),  well above threshold.
\begin{figure}
\mbox{\epsfxsize=10.cm\leavevmode \epsffile{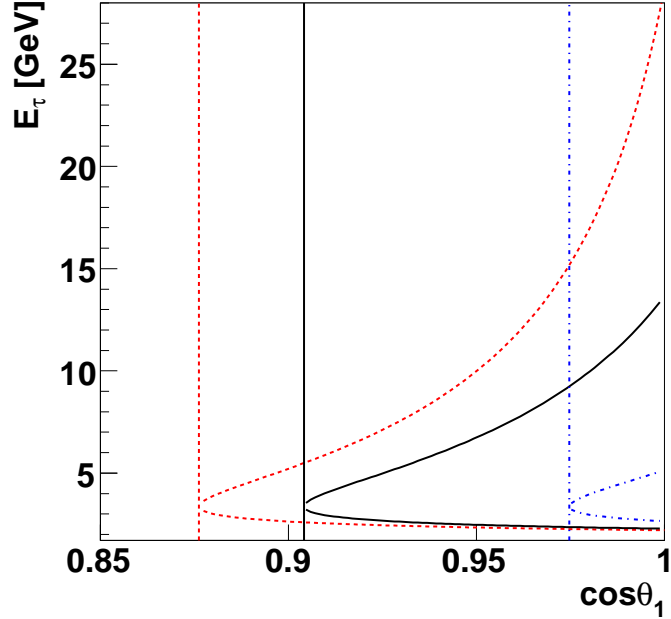}}
%\vspace*{.2 truecm}
\caption{(color online) Energy of the $\tau^-$ lepton as a function of the emission angle for $E=6.85$ GeV (blue dot-dashed line), $E=15$  GeV (black solid line) and $E=30$ GeV (red dashed line), in Lab system. The limiting angles are shown as vertical lines for the corresponding energy.}
\label{Fig:Eth}
\end{figure}

The correlated lepton has two values for the energy (which satisfy energy conservation) and is emitted at two corresponding angles (Fig. \ref{Fig:th12}).  
\begin{figure}
\mbox{\epsfxsize=10.cm\leavevmode \epsffile{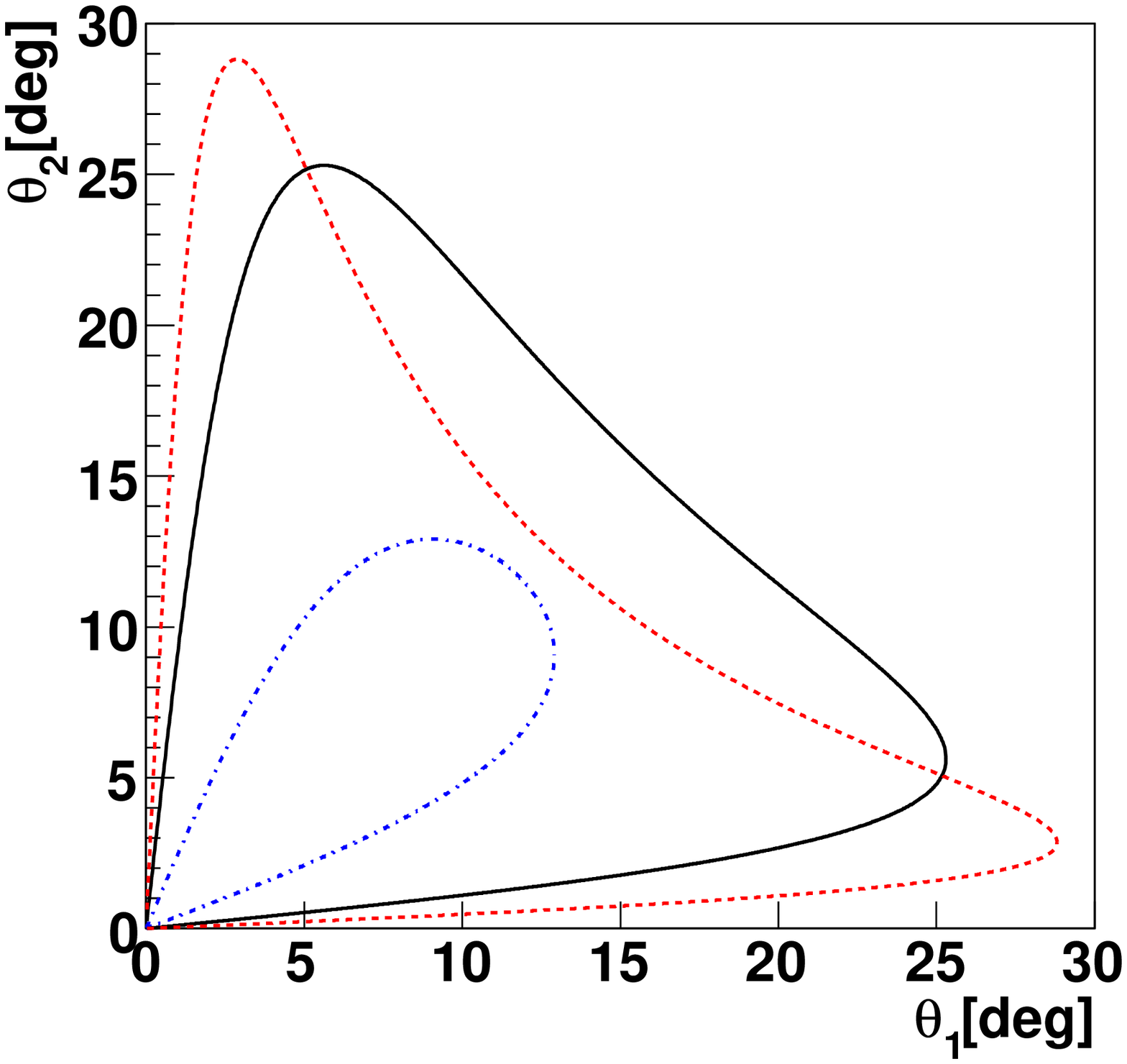}}
%\vspace*{.2 truecm}
\caption{(Color online) $\tau^+$ emission angle $\theta_2$ as a function of the $\tau^-$ angle $\theta_1$ in Lab system. Notations as in Fig. \protect\ref{Fig:Eth}.}
\label{Fig:th12}
\end{figure}

\subsection{The cross section}
%%%%%%%%%%
%%% section results%%%%%%
%%%%%%%%%%%%%%%%%%%%%%%%%%%
In order to illustrate the different polarization observables, in line with previous works (see Ref. \cite{TomasiGustafsson:2005kc}), we choose two parametrizations for time-like FFs. The first one is based on the vector dominance models of Ref. \cite{Ia73}. The second one is a pQCD inspired parametrization, based on analytical extension of the dipole formula in time-like region:
\be
|G_{E,M}^{QCD}|=\frac {A}{(q^2)^2\left [\log^2(q^2/\Lambda^2)+\pi^2 \right ]},~
{A}=96.21~ [\mbox{GeV/c}]^4,
\label{eq:eqqcdbis}
\ee
where $\Lambda=0.3$ GeV is the QCD scale parameter and the value of $A$ has been fitted to the existing data. 

The ratios between the total cross section for an heavy lepton $\ell$ production $\ell=\tau$  ($m_{\tau}$=1776.82 MeV), or $\ell=\mu$ ($m_{\mu}$=105.66 MeV), Eq. (\ref{eq:eqSS}), with respect to the cross section for the production of an electron pair  ($m_e=0.511$ MeV), Eq. (\ref{eq:eqSS0}), is written as:
\be
R_{\ell}=\displaystyle\frac{\sigma (\ell^+\ell^-)}{\sigma (e^+e^-)}=
\displaystyle\frac{1}{2}\beta_{\ell}(3-\beta^2_{\ell}),
\ee
and is illustrated in Fig. \ref{Fig:rap}, as a function of the total energy of the system (from the $\bar p p$ annihilation threshold of $\tau (\mu)$  production, $\sqrt{q^2}=3.5536$ GeV ($\sqrt{q^2}=1.8765$ GeV)). 
The corrections to the ratio due to the mass are of the fourth order and proportional to ($m_{\ell}/\sqrt{q^2})^4$ \cite{Zi62}; therefore, over the kinematical threshold, the  $\mu$ cross section is similar to the electron one. But for $\tau $ production, the variation is significant in the energy region over the $\tau^+\tau^- $ threshold. 

\begin{figure}
\mbox{\epsfxsize=10.cm\leavevmode \epsffile{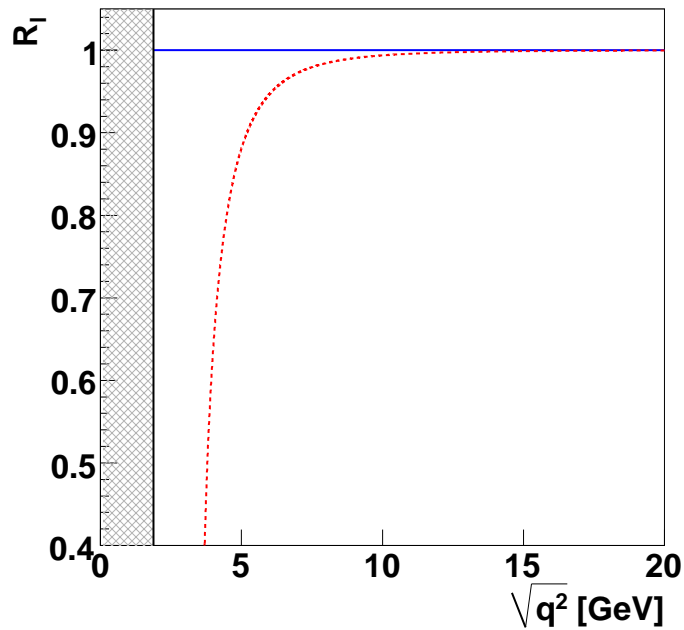}}
%\vspace*{.2 truecm}
\caption{(Color online) Total cross section ratios  $R_{\ell}=\frac{\sigma (\ell^+\ell^-)}{\sigma (e^+e^-)}$,
for $\ell=\tau$ (red dashed line) and  $\ell=\mu $ (blue solid line) as function of $\sqrt{q^2}$. The shaded area illustrates the region below the physical threshold for $\bar p+p$ annihilation. }
\label{Fig:rap}
\end{figure}

The angular dependence of the differential cross section is shown in Fig. \ref{Fig:csec}, at a fixed value of $q^2$. For illustration, we use here the FFs parametrization from \cite{Ia73}. 

As mentioned above, the effect of the mass is to change the stiffness of the angular dependence of the differential cross section as function of $\cos\theta$, as illustrated in Fig. \ref{Fig:csec} (left), for $q^2=15$ GeV$^2$. For $\tau$ the relative contribution of the electric to magnetic term is larger. In other words, the effect of the mass is to change the slope and the intercept of the linear dependence of the differential cross section as function of $\cos^2\theta$, as illustrated in Fig. \ref{Fig:csec} (right).

\begin{figure}
\mbox{\epsfxsize=10.cm\leavevmode \epsffile{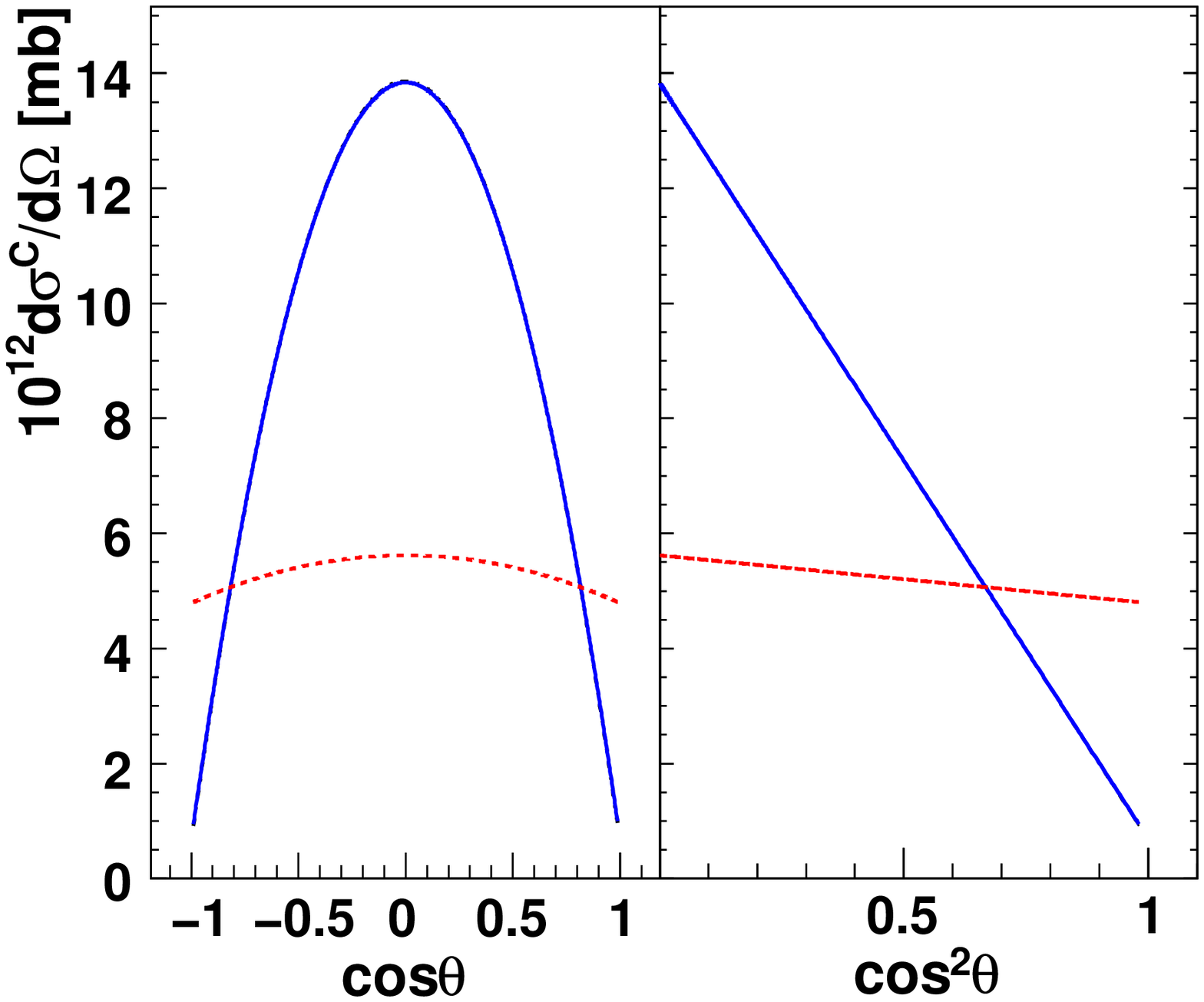}}
%\vspace*{.2 truecm}
\caption{(color online)Differential cross section  as a function of $\cos\theta$ (left);  as a function of $\cos^2\theta$ (right) for $q^2$=15 GeV$^2$, assuming the parametrization \protect\cite{Ia73}, for $\ell=\tau$ (red dashed line), $\ell=\mu $ (blue solid line). The calcualtion for $\ell =e $ (black dotted line) is hardly visible since it overlaps with the $\mu$ line.}
\label{Fig:csec}
\end{figure}

\subsection{Polarization observables}

The CMS $q^2$ dependence of the single spin asymmetry, as well of the triple polarization observables and of some double spin observables, is driven by the term $Im G_MG_E^*$ (besides simple kinematical coefficients), therefore it will constitute a direct test of nucleon models. The single spin asymmetry has been calculated for $e$, $\mu$ and $\tau$ lepton pair production, at $q^2=15 $ GeV$^2$,  for the FF parametrization ot Ref. \cite{Ia73} in CMS (Fig. \ref{Fig:sasym}). It is very small for $\tau$ lepton, whereas for $\mu$ and $e$ it exhibits a strong forward backward angular asymmetry. In the energy distribution (Lab system) a pronounced structure is visible around $E\sim 7$ GeV, at $\cos\theta_1=0.95^{\circ}$ for $\mu$ and $e$ (Fig. \ref{Fig:asymlab}). For $\tau$ the this observable is small, in all the energy range. The double solution is shown.

\begin{figure}
\mbox{\epsfxsize=10.cm\leavevmode \epsffile{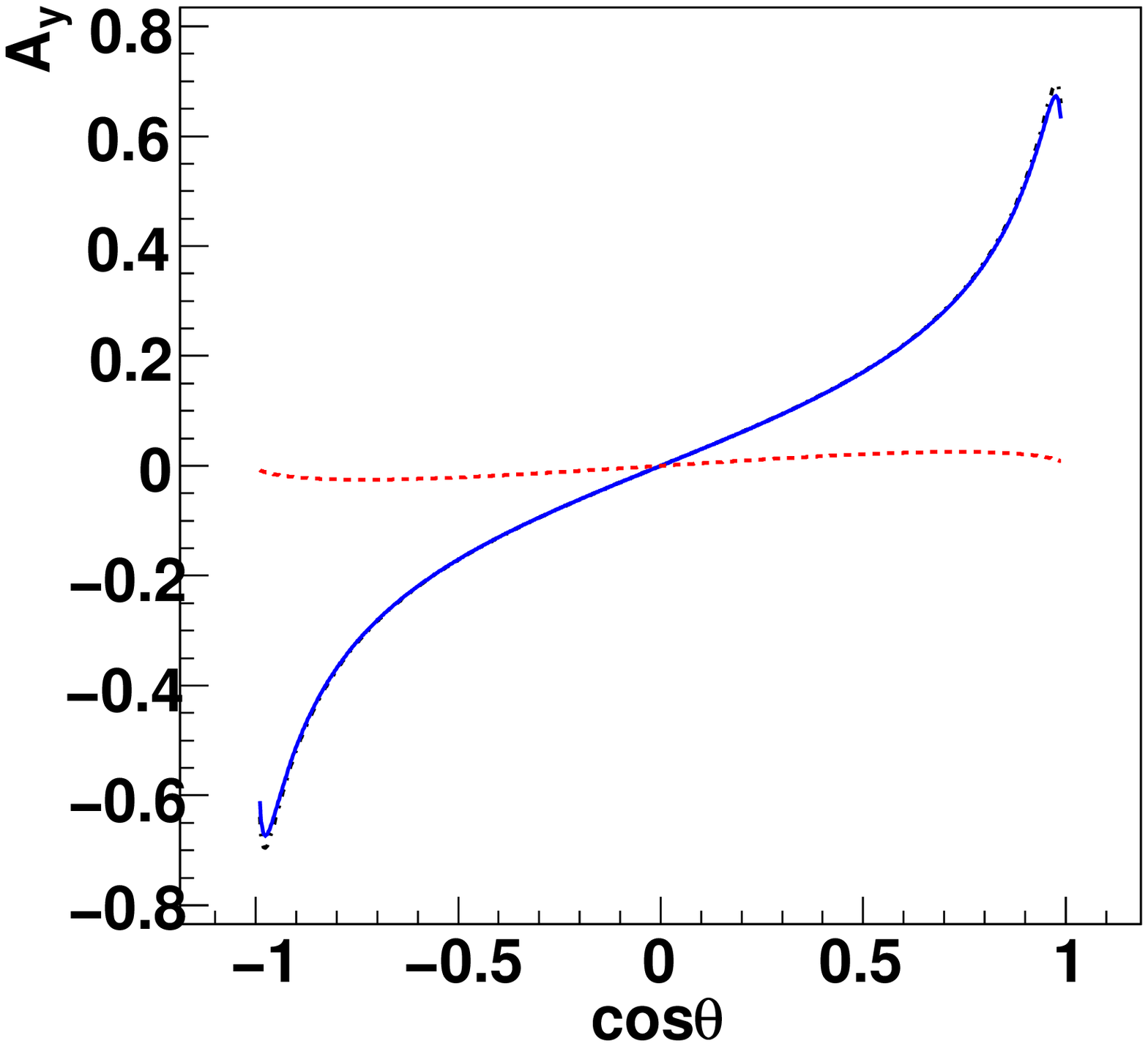}}
%\vspace*{.2 truecm}
\caption{(color online) Single spin asymmetry as a function of $\cos\theta$  for $q^2$=15 GeV$^2$, for the parametrization \protect\cite{Ia73}.  Notations as in Fig. \protect\ref{Fig:csec}.}
\label{Fig:sasym}
\end{figure}

\begin{figure}
\mbox{\epsfxsize=10.cm\leavevmode \epsffile{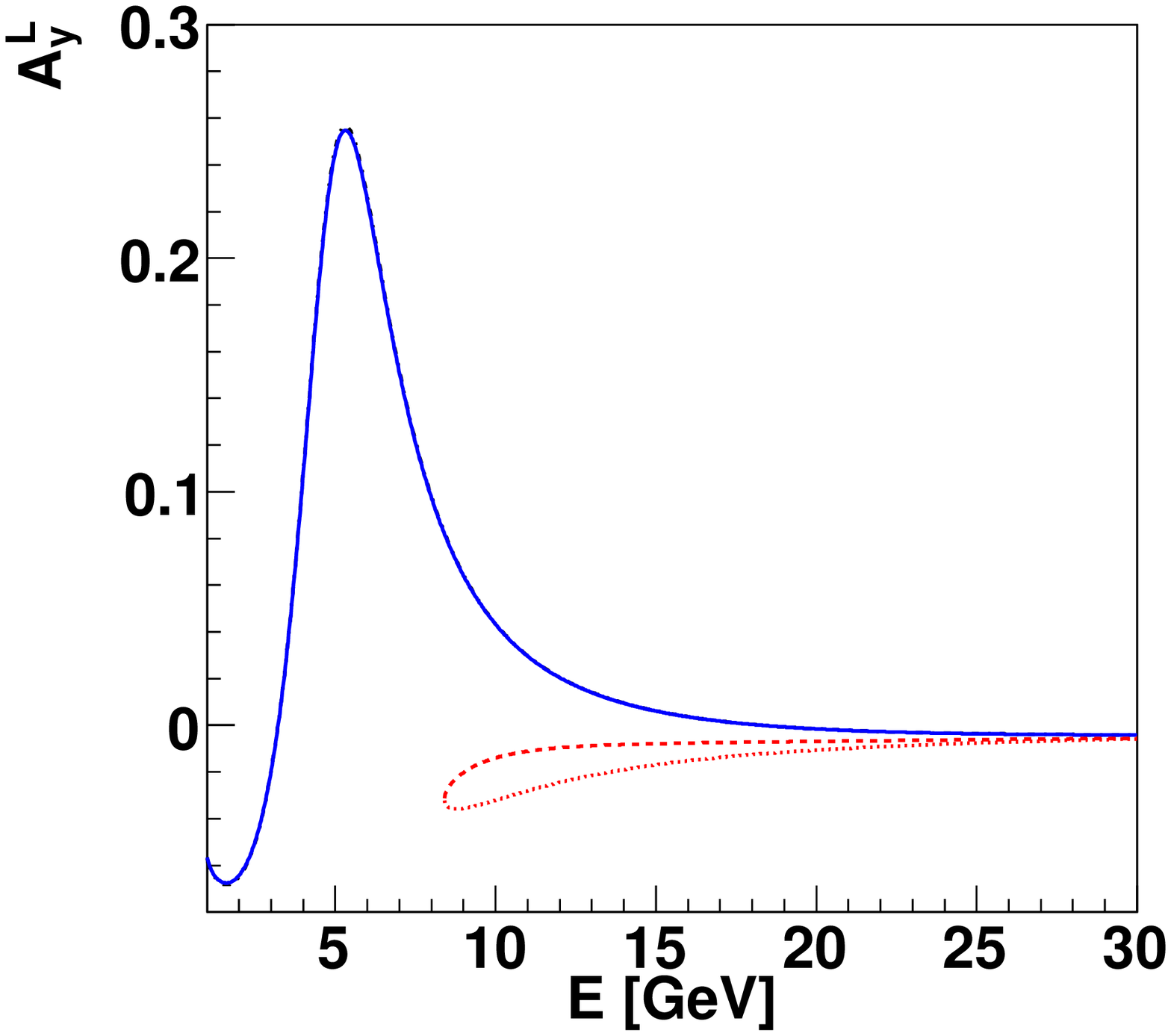}}
%\vspace*{.2 truecm}
\caption{(color online) Single spin asymmetry as a function of the antiproton energy in Lab system, for $\cos\theta_1=0.95$, for the parametrization \protect\cite{Ia73}, for $\mu$ (blue solid line), for $e$ (black dash-dotted line), and for $\tau$. corresponding to $E_1^+$ (red dashed line) and to $E_1^-$ (red dotted line).}
\label{Fig:asymlab}
\end{figure}

The non vanishing double spin observables are shown in Fig. \ref{Fig:Double} for the parametrization of Ref. \cite{Ia73} as a function of $\cos\theta$ in CMS for $E=15$ GeV. From top to bottom, from left to right are illustrated:  the polarization transfer coefficients,  $T_{\ell x}^C$, $T_{\ell z}^C$, $T_{n y}^C$, $T_{t x}^C$, $T_{t z}^C$ from Eqs.(\ref{eq:eqt}), 
the  analyzing powers in polarized proton-antiproton collisions $ A^C_{xx}$, $A^C_{yy}$, $ A^C_{zz}$, $ A^C_{xz}$ 
from  Eqs. (\ref {eq:e4}), and the correlation coefficients when the polarization of the lepton-antilepton pair is measured: 
$ C^C_{nn}$, $C^C_{tt}$, $ C^C_{\ell\ell}$, $ C^C_{\ell t}$  
from  Eqs.(\ref{eq:eqC}). Note that $A^C_{yy}$ coincides with $C^C_{nn}$ and it is not shown.

In Fig.  \ref{Fig:pqcd}, the observables are shown when FFs are calculated from Eq. \ref{eq:eqqcdbis}, where $|G_E|=|G_M|$. Note that this pQCD inspired parametrization does not have any imaginary part, therefore some of the  observables vanish. From top to bottom, from left to right are illustrated:  the polarization transfer coefficients,  $T_{\ell z}^C$, $T_{t z}^C$ from  Eqs. (\ref{eq:eqt}), the  analyzing powers in polarized proton-antiproton collisions $ A^C_{xx}$, $A^C_{yy}$, $ A^C_{zz}$ 
from  Eqs. (\ref {eq:e4}), and the correlation coefficients when the polarization of the lepton-antilepton pair is measured: 
$ C^C_{nn}$, $C^C_{tt}$, $ C^C_{\ell\ell}$, $ C^C_{\ell t}$ 
from  Eqs. (\ref{eq:eqC}). 

For the $\tau$-meson, the effect of the mass is sizable in  all the observables. The difference between $\mu$ and $e$ is tiny and it is best seen in the observables related to the transverse polarization, such as $T_{tz}$ and $C_{\ell t}$. This effect is relatively larger when the incident energy is smaller. 
\begin{figure}
\mbox{\epsfxsize=15.cm\leavevmode \epsffile{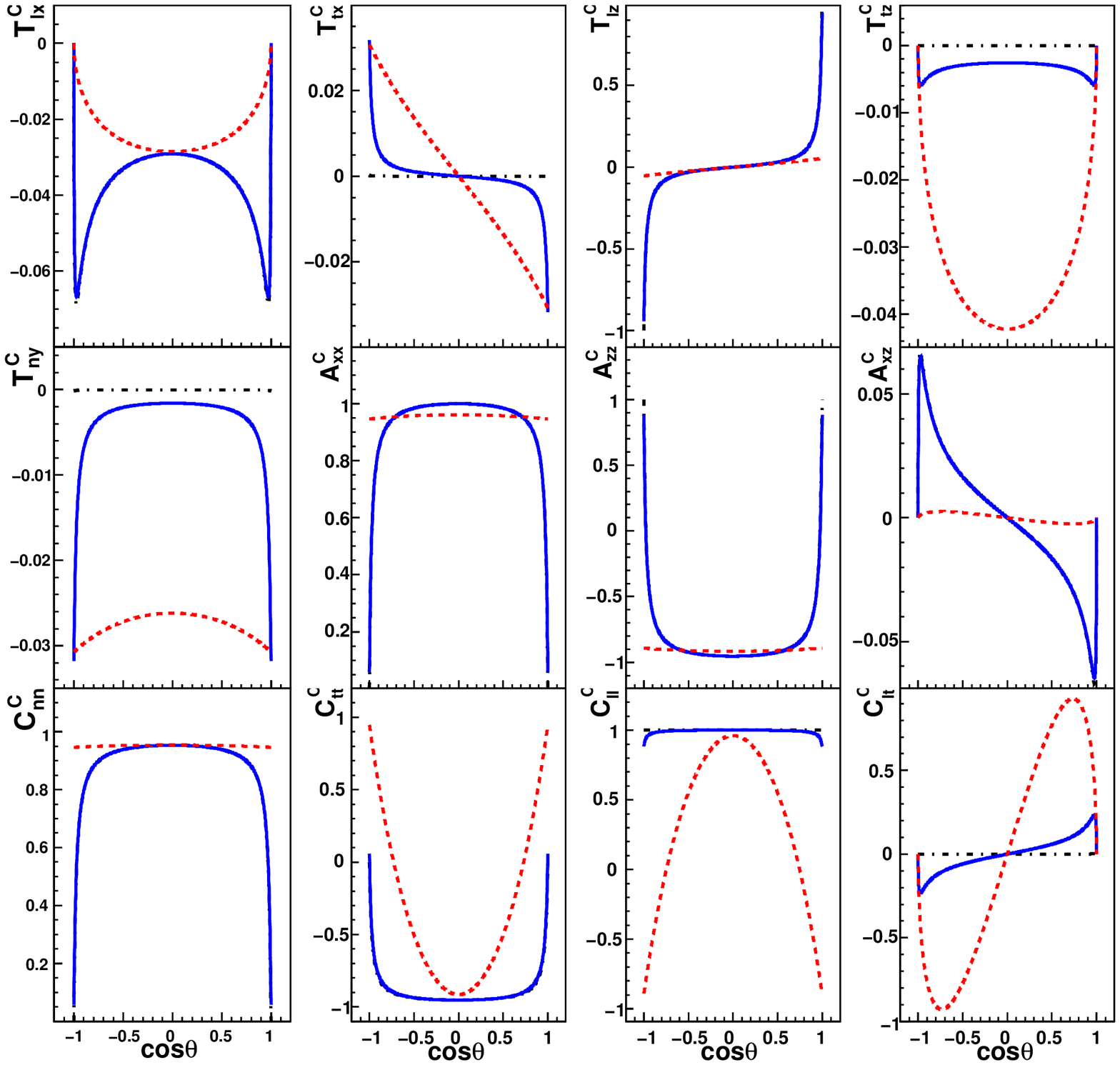}}
%\vspace*{.2 truecm}
\caption{(color online) Double polarization observables  as a function of $\cos\theta$ , for $q^2=15$ GeV$^2$, using  the parametrization from Ref. \protect\cite{Ia73} in CMS. Notations as in Fig. \protect\ref{Fig:csec}.}
\label{Fig:Double}
\end{figure}
\begin{figure}
\mbox{\epsfxsize=15.cm\leavevmode \epsffile{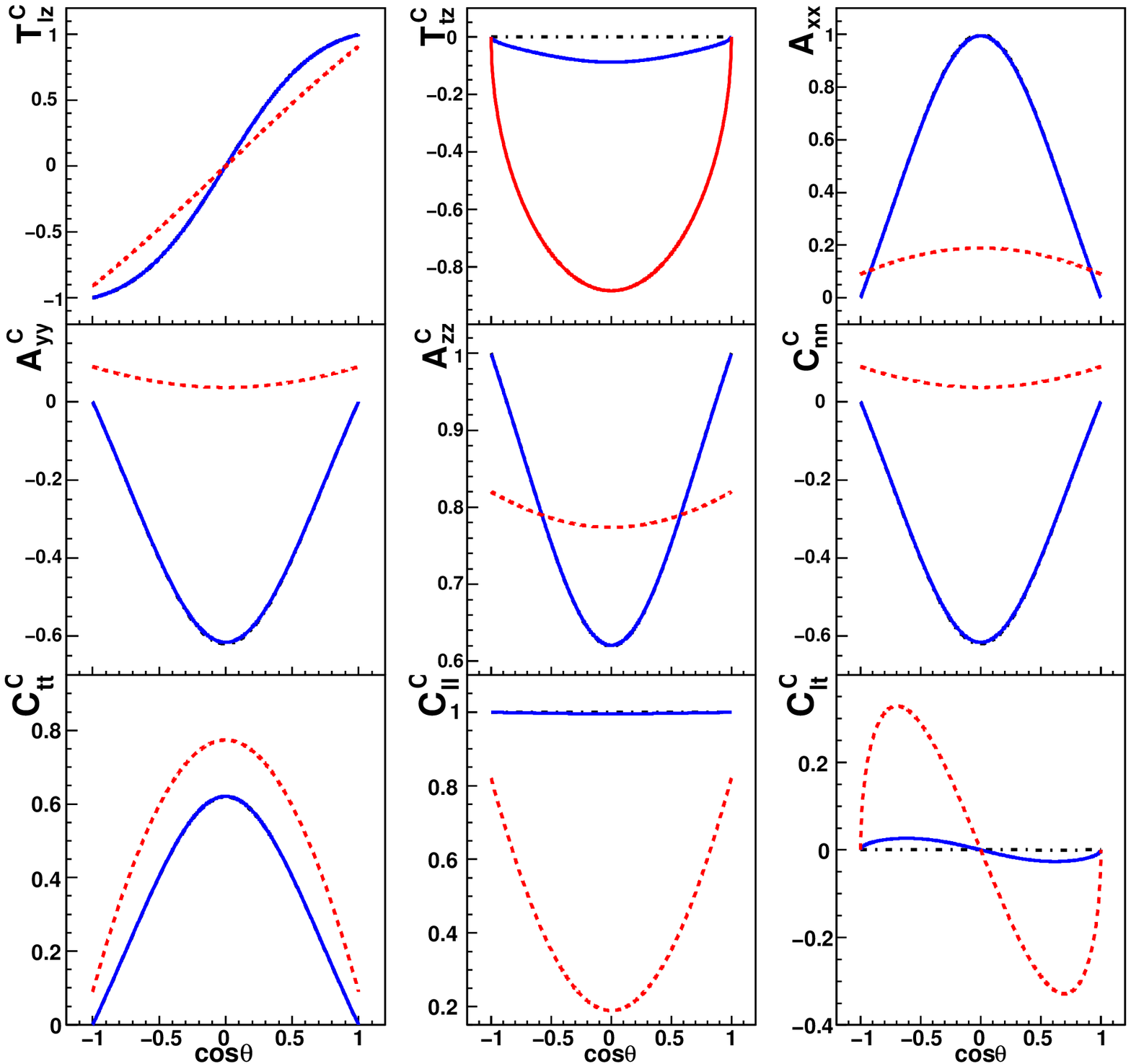}}
%\vspace*{.2 truecm}
\caption{(color online) Double polarization observables as a function of $\cos\theta$ , for $q^2=15$ GeV$^2$, using the parametrization of FFs from Eq. (\protect\ref{eq:eqqcdbis}) in CMS. Notations as in Fig. \protect\ref{Fig:csec}.}
\label{Fig:pqcd}
\end{figure}

\section{Conclusions}
The calculation of polarization observables for the annihilation of proton-antiproton into a lepton pair was extended to the case of heavy leptons, such as $\tau$. In this case it is not possible to  neglect the lepton mass. The calculation was performed in the one-photon exchange approximation, The expressions of the observables are given in terms of nucleon electromagnetic FFs.

We calculated the polarization observables in the reaction CMS, which is considered as the natural frame for the study of annihilation reactions, and also in the Lab frame, as this reaction may be studied in principle, at the PANDA experiment which is a fixed target experiment.

The following cases were considered: the antiproton beam or the proton target are polarized (the one single beam or target asymmetry), the antiproton beam and the final lepton are polarized (the polarization transfer from the antiproton beam to the detected lepton), the antiproton beam and the proton target
are both polarized (the double spin asymmetry coefficients), and finally when both leptons are polarized (the correlation coefficients). We also gave expressions for some triple polarization observables: when both hadrons in  initial state and one final lepton are polarized, and in the case when the antiproton beam is polarized and the polarization of both leptons is measured.

We investigated the dependence of the unpolarized cross section, of the angular asymmetry and various polarization observables on the mass of the lepton. It was found that at small angles, the contribution which is proportional to $|G_E|^2$ dominates in the double analyzing powers $A_{xx}^C$ and $A_{yy}^C$ (the case when both antiproton beam and proton target are polarized) and this effect arises due to the heavy lepton mass. In all other observables the electric contribution is suppressed by the factor $\eta_p^{-1}$.

%%%%%%%%%%%%%%%%%%%%%%%%%%
\section{Acknowledgments}
%%%%%%%%%%%%%%%%%%%%%%%%%%
One of us (A.D.) acknowledges the Libanese CNRS for financial support. This work was partly supported by  CNRS-IN2P3 (France) and by the National Academy of Sciences of Ukraine under PICS No. 5419 and by GDR No.3034 "Physique du Nucl\'eon" (France). We acknowlege E. A. Kuraev, Th. Hennino, J. Van de Wiele,  for interesting discussions, F. Maas and Ch. Morales for interest in our work.

%%%%%%%%%%%%%%%%%%%%%%%%%%%%%%%%%%%%%%%%%%

\end{document}